\documentclass[]{pasj01}
\draft
\usepackage{comment}

\begin{document} 
\Received{}
\Accepted{}

\title{Discovery of a keV-X-ray Excess in RX J1856.5--3754}

\author{Tomokage \textsc{yoneyama}\altaffilmark{1,2}%
}
\altaffiltext{1}{Department of Earth and Space Science, Graduate School of Science, Osaka University, 1-1 Machikaneyama-cho, Toyonaka, Osaka 560-0043, Japan}
\altaffiltext{2}{Project Research Center for Fundamental Sciences, Graduate School of Science, Osaka University, 1-1 Machikaneyama-cho, Toyonaka, Osaka 560-0043, Japan}

\email{yoneyama@ess.sci.osaka-u.ac.jp}

\author{Kiyoshi \textsc{hayashida}\altaffilmark{1,2}}

\author{Hiroshi \textsc{nakajima}\altaffilmark{1,2}}

\author{Shota \textsc{inoue}\altaffilmark{1,2}}

\author{Hiroshi \textsc{tsunemi}\altaffilmark{1,2}}


\maketitle

\begin{abstract}
RX J1856.5$-$3754 is the brightest and nearest ($\sim 120$ pc) source among thermally emitting isolated neutron stars. Its spectra observed with {\sl XMM-Newton} and {\sl Chandra} satellites are well-fitted with the two-temperature  ($kT^\infty \sim$ 32 and 63 eV) blackbody model. Fitting ten sets of the data from {\sl Suzaku} XIS0, XIS1, XIS3 and {\sl XMM-Newton} EPIC-pn with the two-temperature blackbody model, we discover an excess emission, 16--26\% in 0.8--1.2\,keV. We examine possible causes of this keV-X-ray excess; uncertainty in the background, pile up of the low energy photons and confusion of other sources. None of them succeeds in explaining the keV-X-ray excess observed with different instruments. We thus consider this keV-X-ray excess is most likely originated in RX J1856.5$-$3754. However, it is difficult to constrain the spectral shape of the keV-X-ray excess. The third blackbody component with $kT^\infty = 137^{+18}_{-14}$ eV, additional power-law component with a photon index $\Gamma = 3.4^{+0.5}_{-0.6}$, or Comptonization of blackbody seed photons into power-law with a photon index $\Gamma_c = 4.3^{+0.8}_{-0.8}$ can reproduce the keV-X-ray excess. We also search for the periodicity of 0.8--1.2\,keV data, since 7.055 s pulsation is discovered from 0.15--1.2\,keV band in the XMM Newton EPIC-pn data ($\sim$1.5\%). We only obtain the upper limit of pulsed fraction $<$ 3\% in the keV-X-ray excess. 
We shortly discuss the possible origin of the keV-X-ray excess, such as synchrotron radiation and Comptonization of blackbody photons.
\end{abstract}

\section{Introduction}

ROSAT soft X-ray survey and pointing observations have discovered seven isolated radio-quiet neutron stars showing blackbody-like X-ray spectra with $kT^\infty \sim$ 50 -- 100\,eV (e.g. \cite{trumper_2004}, \cite{haberl_2007}), where $kT^\infty$ is the apparent blackbody temperature observed in which gravitational redshift is not corrected. Small amount of interstellar absorption observed in the X-ray spectra suggests that these sources are nearby objects having X-ray luminosities $L_X$ of $10^{30}$−$10^{32}$ erg s$^{-1}$ (\cite{trumper_2004}), and thus called X-ray dim isolated neutron stars (XDINSs). They are the targets of extensive studies as promising objects to learn about the atmosphere and internal structure of neutron stars (e.g. \cite{paerels_1997}), and the equation of state of neutron star matter (e.g. Pons et al. 2002, \cite{trumper_2004}). 

The first identification of XDINSs was done by Walter et al. (1996) for RX J1856.5$-$3754 (J1856 hereafter), which is the brightest among seven sources. Parallax measurements employing the Advanced Camera for Surveys (ACS) on the Hubble Space Telescope (HST) determined the distance to J1856 to be 123$^{+11}_{-15}$ pc (\cite{walter_2010}). Observations of this source were repeated with X-ray satellites after ROSAT (e.g. \cite{burwitz_2003}, \cite{beuermann_2006}, \cite{sartore_2012}) not only for science purpose but also for calibration purpose. The latter is due to the fact that the X-ray spectrum of J1856 is ideal for instrumental calibration in soft X-ray range (e.g. 0.2--1.0\,keV) and is (or believed to be) stable over years. Burwitz et al. (2001) studied the X-ray spectra of this source obtained with Chandra/LETG. As following study, Burwitz et al. (2003) analyzed the data obtained with Chandra/LETG and XMM-Newton/EPIC and RGS. In both studies, all the spectra are consistently fitted with a simple blackbody model with $kT^\infty \sim $ 63 eV, and $N_H \sim 1 \times 10^{20}$  cm$^{-2}$ with no significant line features, while some spectral features have been reported for other XDINSs (\cite{haberl_2007}, \cite{hohle_2012}). Evaluated radius of the emission region ($R^\infty$) is $\sim$ 4.4 km at 120 pc. This is too small for a neutron star, and in fact, a possibility of a quark star was claimed by authors (e.g. \cite{drake_2002}, \cite{kohri_2003}).

Burwitz et al. (2001 and 2003) also argued various neutron star atmosphere models (see \cite{zavlin_2002}). They mentioned that the previous study of light elements (H and He) atmosphere model (Pavlov et al. 1996 and Pons at al. 2002) could not reproduce the spectra obtained with ROSAT/PSPC. Then they applied the Fe atmosphere or the solar-mixture atmosphere model as alternatives against the blackbody model, but concluded that they are unable to reproduce the X-ray emission of J1856, while a condensed matter surface (e.g. \cite{lai_2001}) might be a possibility. Ho et al. (2007) suggest that their partially ionized hydrogen atmosphere model can also reproduce the spectrum, which is almost the same with the blackbody model in X-ray band.

On the other hand, a simple blackbody model fit leads to another problem, i.e., the flux of this source in the optical band is about a factor of 7 larger than that given by the extension of the X-ray data. Pons et al. (2002) introduced a two-temperature (2T) blackbody model in their analysis of the ROSAT data. Burwitz et al. (2003) applied this model to their data, providing the hot component parameters, $kT_h^\infty \sim$ 64 eV and $R_h^\infty$ = 4.4 km at 120 pc for the X-ray emission region, with the cool component parameters of $kT_c^\infty < 33$ eV and $R_c^\infty > 17$ km. Beuermann et al. (2006) applied the same model to the Chandra LETG+HRC data and provided $kT_h^\infty = 62.83\pm0.41$ eV and $R_h^\infty = 4.54\pm0.03$ km with $kT_c^\infty = 32.26\pm0.72$ eV and $R_c^\infty = 16.5\pm0.12$ km, where $N_H =(11.0\pm0.3) \times 10^{19}$ cm$^{-2}$. These parameters have been often used for soft X-ray calibration of the instruments. Sartore et al. (2012) employed the XMM-Newton EPIC-pn data of this source obtained in the interval from April 2002 to October 2011 to search for a long-term variability. They found only small variations of the X-ray spectra in 0.15-1.2 keV range, less than 1\% in terms of the blackbody temperature. They deduced these variations were partly due to instrumental effects. They also provided the spectral parameters of the 2T blackbody model; $N_H = 12.9^{+2.2}_{-2.3} \times 10^{19}$ cm$^{-2}$, $kT^\infty_h  = 62.4^{+0.6}_{-0.4}$ eV, $R^\infty_h = 4.7^{+0.2}_{-0.3}$ km, $kT^\infty_s  = 38.9^{+4.9}_{-2.9}$ eV and $R^\infty_s = 11.8^{+5.0}_{-0.4}$ km. 

When we adopt a 2T blackbody model for J1856, a hot component may correspond to the hot spot on the surface of the neutron star. In this case, pulsations in the X-ray light curve should be observed. After extensive search for the pulsation from this source, Tiengo and Mereghetti (2007) finally detected pulsations with a period of 7.055 s in the XMM-Newton observations. Their pulsed fraction in the 0.15--1.2\,keV band is about 1.2\%, the smallest ever observed in XDINSs. They analyzed the phase-resolved spectra, but obtained no significant evolution of the blackbody temperature and only obtained an upper limit of 1.2 eV at 90\% confidence level as the temperature difference of the blackbody. They also derived an upper limit of the period derivative $\dot{P} < 1.9 \times 10^{-12}$ s s$^{-1}$. This corresponds to the magnetic field $B < 1.2 \times 10^{14}$ G, under the usual assumption of vacuum dipole magnetic breaking. Karkwijk \& Kaplan (2008) estimated $\dot{P} \sim 2.97 \times 10^{-14}$ s s$^{-1}$, corresponds to $B \sim 1.5 \times 10^{13}$ G employing the XMM-Newton EPIC and Chandra HRC data with the coherent analysis. 

Although the origin of the hot component for J1856 is still an open question, the recent discovery of the vacuum birefringence, which is the first evidence of quantum electric dynamical effects in the strong magnetic field, from the optical observations for J1856 (Mignani et al. 2017) encourages the further studies of the X-ray observations. This source also becomes one of the most important calibration targets for soft X-ray instruments on X-ray astronomy satellites. In fact, the Suzaku satellite (\cite{mitsuda_2007}) observed this source 17 times in its lifetime, primarily for the calibrations of the X-ray Imaging Spectrometers (XIS) (\cite{koyama_2007}). The data obtained in these observations have been used to make empirical models of contaminants accumulated on the optical blocking filters (OBFs) of the XIS cameras (Suzaku technical description\footnote{http://www.astro.isas.ac.jp/suzaku/doc/suzaku\underline{ }td}). In the course of this effort, we noticed a systematic excess of the X-ray counts from the source around and above 1 keV, where the contaminant can little affect the efficiency. As the X-ray spectrum of J1856 is extremely soft, authors so far have been focusing on the energy range below 1 keV and paid little attention above 1 keV. 

In this paper, we examine a possible hard excess over the 2T blackbody model for J1856. We first show such an excess in the data obtained with the Suzaku XIS1 (BI CCD) of which soft X-ray efficiency is higher than other XIS CCD (FI CCD). One of the points of our investigations is to evaluate various possible systematic uncertainties in the data reduction. In particular, as the target of our study, J1856, is also used for calibration, we need special caution on those uncertainties. We thus employ other two XIS cameras on Suzaku and also the EPIC-pn on XMM-Newton. 

In section 2, we describe the observations, data reduction and the X-ray spectral analysis using the 2T blackbody model provided by Beuermann et al. (2006) based on the XMM-Newton EPIC-pn and to see a possible excess around 1 keV. In section 3, we quantitate it and investigate various possibilities that produce such an excess and evaluate the systematic uncertainties. We try to model the keV-excess emission in section 4, a timing analysis in section 5 and discuss possible origins in section 6. 

Unless otherwise noted, the errors of the parameters written in the text and tables are those with 90\% confidence level. The error bars in the plots are those with 68\% confidence level.

\section{Data reduction and spectral fit with the 2T blackbody model}

We use the data observed by Suzaku from 2005 to 2014, listed in table \ref{tab:suzaku_data}.
From each observation, we employ cleaned event data of the XIS. We extract events from the circular region within a 2$^\prime$.17 radius around the target centroid to make a source spectrum, and background spectrum is obtained from 2$^\prime$.90--5$^\prime$.79 annulus region around the source. This process is performed with \verb'XSELECT V2.4d'. Considering the fact that the sky background is affected by a telescope vignetting effect, we apply the vignetting correction to the background. We evaluate the correction factor by making an exposure map with \verb'xissim', assuming the background in the soft X-ray band is dominated by the sky background.  Thus we obtain a vignetting-corrected background with multiplying the background count rate by 1.184. The extracted spectra are binned in order to have more than 30 counts per bin using \verb'grppha'. Redistribution matrix files (RMFs) and auxiliary response files (ARFs) are created with \verb'xisrmfgen' and \verb'xissimarfgen' in the HEASOFT 6.19. The spectral analysis is performed with \verb'XSPEC ver12.9.0n', selecting the energy range in 0.25-2.0 keV.

\begin{figure}
 \begin{center}
  \includegraphics[width=10cm]{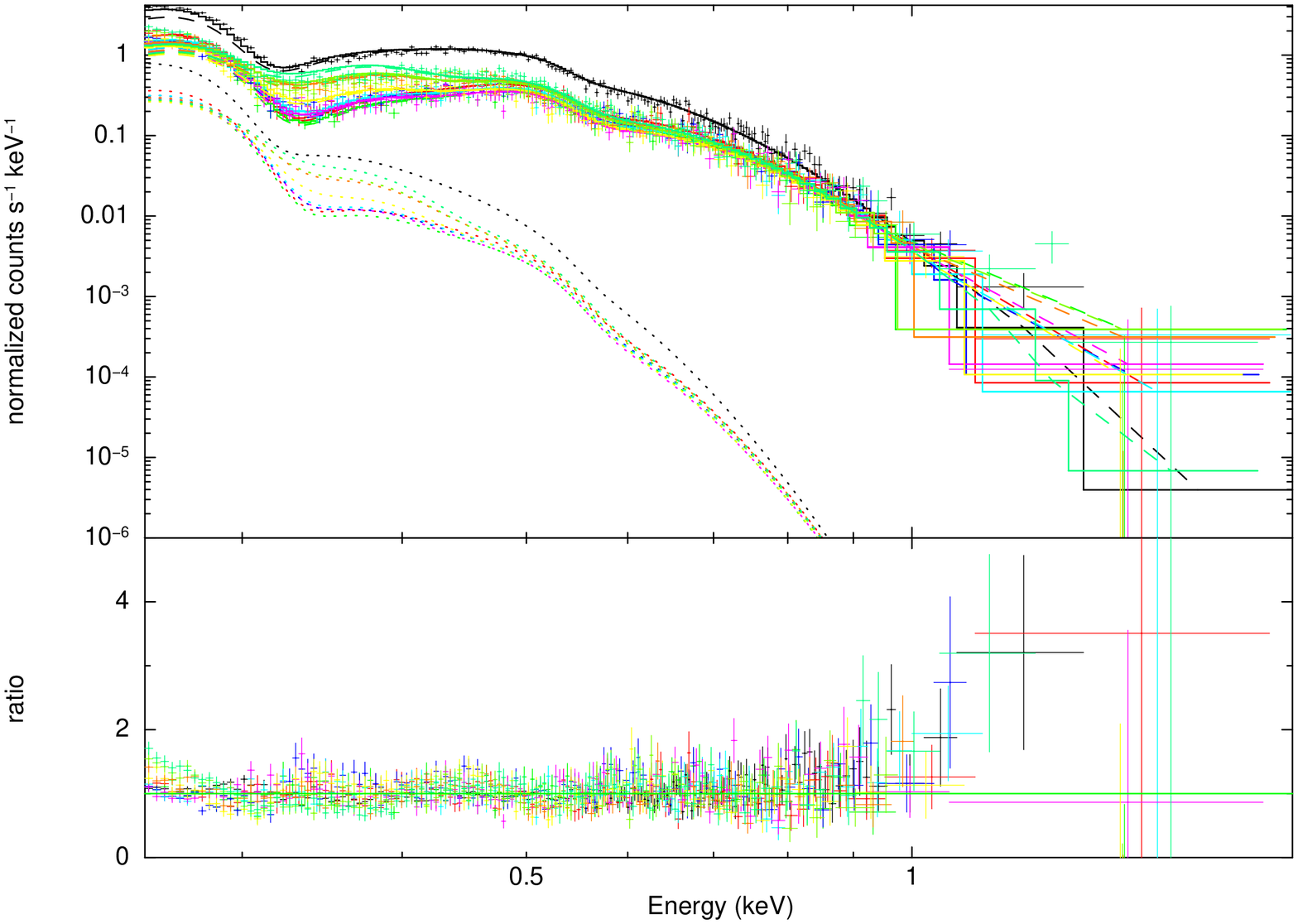} 
 \end{center}
\caption{Background-subtracted spectra of J1856 observed with Suzaku XIS1 during 2005--2014. The solid lines are 2T blackbody model for each observation, whose parameters are fixed to those obtained by Beuermann et al. (2006) with $\chi^2_r = 1.88$ for 1307 dof. The dashed lines are the hot component and the doted lines are the cold component. We find that the data exceed the model in the $> 0.8$ keV energy range.}\label{fig:2bb_era}
\end{figure}

We first fit the spectra collected by XIS1 with the 2T blackbody model, schematically, \verb'const*phabs*(bbodyrad+bbodyrad)', whose parameters are fixed to the best fit values of Beuermann et al. (2006). These are often used for the calibration of various soft X-ray instruments. We employ \verb'const' as a free parameter for each spectrum. The best fit values of \verb'const' range from 1.1 to 1.2, and we obtain a reduced $\chi^2$ value, $\chi^2_r =$ 1.88 for 1307 degrees of freedom (hereafter dof). As shown in figure \ref{fig:2bb_era}, we notice the data exceed the model nearly twice in the $> 0.8$ keV energy range. We then merge all the spectra whose  total exposure is 450.4 ks and fit them with the same model. The best fit results gives a constant factor $1.13 \pm 0.004$ and $\chi^2_r =$ 2.73 for 230 dof. We find the data exceed the model by a factor $\sim 1.7$. Hereafter we call this ``keV-X-ray excess'' or ``keV-excess''.

\begin{figure}
 \begin{center}
  \includegraphics[width=10cm]{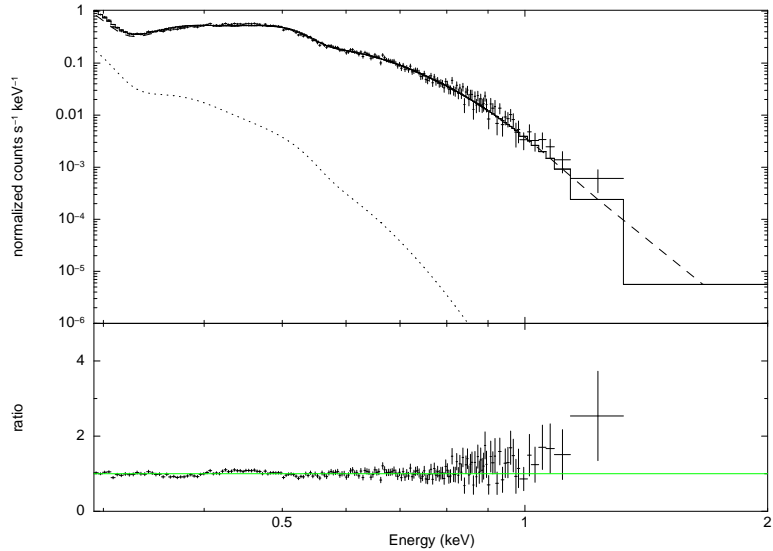} 
 \end{center}
\caption{XIS1 merged spectrum fitted with the 2T blackbody model. Correction to the OBF contamination model is taken into account with an absorption model of H,C, and O. Although we obtain better fit with $\chi^2_r = 1.86$ for 230 dof, the keV-excess is visible.}
\label{fig:2bbhco_sum}
\end{figure}

We notice that the fit is not good below 0.7 keV. We think it is due to systematic error in the calibration of the OBF contamination, though the contamination is modeled and taken into account in the ARFs of each epoch. We thus introduce an absorption model of hydrogen, carbon, and oxygen (dominant components of the contaminant) to compensate the error. 
We obtain better fit with $\chi^2_r =$ 1.86 for 230 dof as shown in figure \ref{fig:2bbhco_sum}. However, even in this case, the keV-excess still remains.

\begin{figure}
 \begin{center}
  \includegraphics[width=10cm]{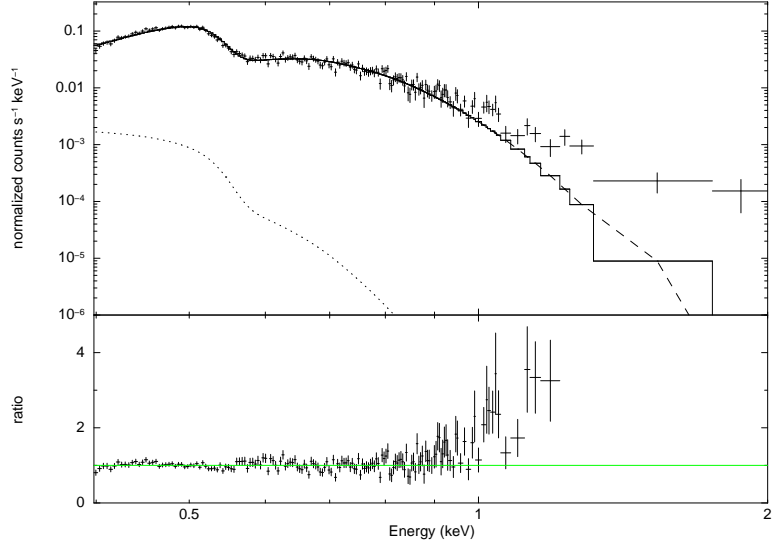} 
 \end{center}
\caption{Spectra of XIS FI CCDs with the contamination-corrected 2T blackbody model.The fit provides $\chi^2_r = 1.40$ for 187 dof, but the keV-excess is visible in $>0.8$ keV energy range.}
\label{fig:2bbhco_03}
\end{figure}

\begin{figure}
 \begin{center}
  \includegraphics[width=10cm]{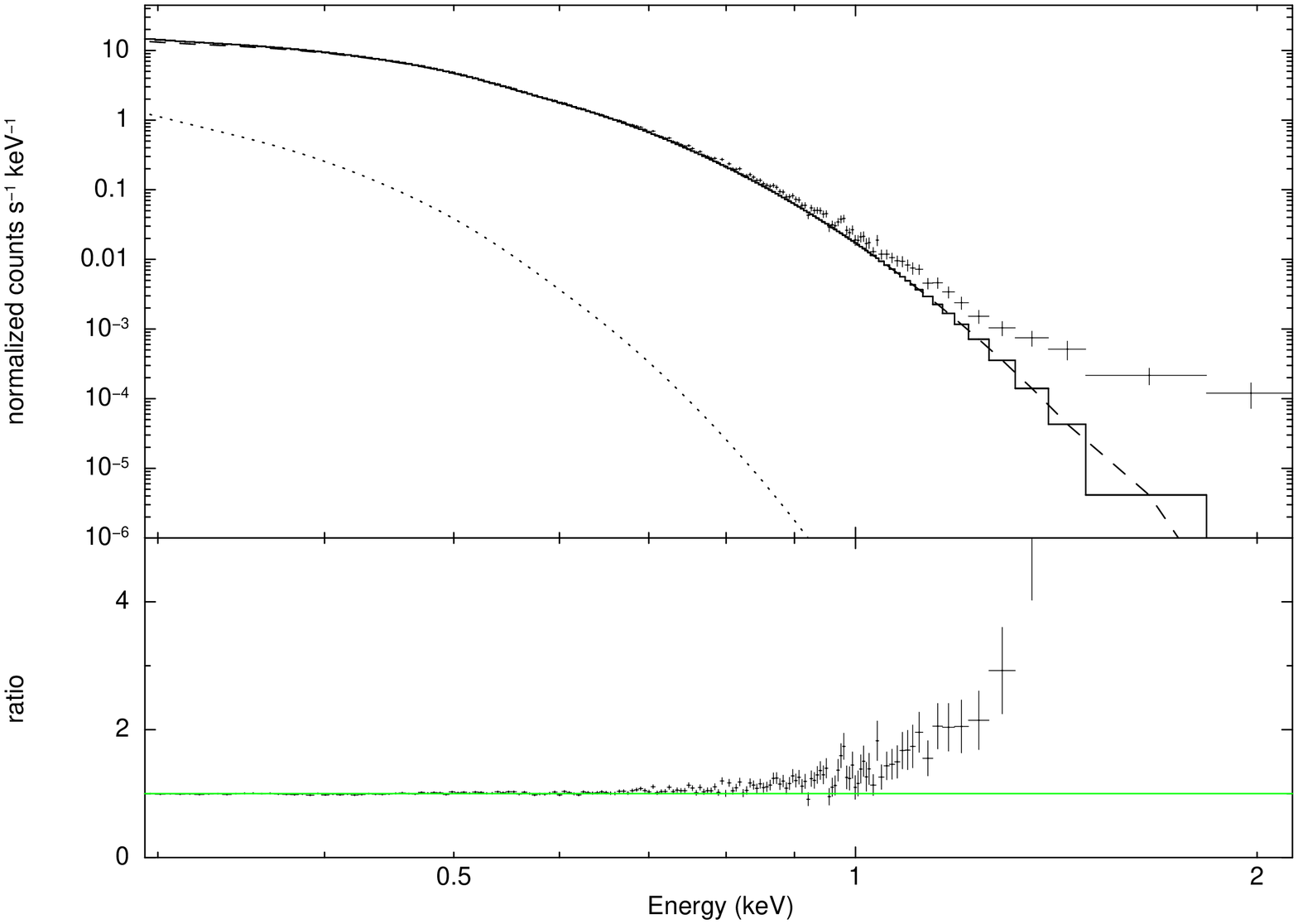} 
 \end{center}
\caption{Merged spectrum of the EPIC-pn data listed in table \ref{tab:xmmdata}, obtained in twelve observations during 2004--2015 with the 2T blackbody model which gives $\chi^2_r = 3.06$ for 147 dof. Similarly the keV-excess is apparent.}\label{fig:epn_sum}
\end{figure}

To examine the keV-excess, we also analyze data obtained by other instruments, XIS0, XIS3 onboard Suzaku.
For XIS0 and XIS3, we merge together all of their spectra because both of them are FI CCD. We follow the same reduction procedure as that for XIS1, but select 0.4--2.0\,keV range. The result is shown as figure \ref{fig:2bbhco_03}, with $\chi^2_r = $1.40 for 187 dof.
We also employ the data of Europian Photon Imaging Camera pn (EPIC-pn) onboard XMM-Newton, listed in table \ref{tab:xmmdata}. All of these observations are performed in small window mode with thin filter. We process raw event data with \verb'evselect' of the XMM-Newton science analysis package (SAS) v14.00, selecting only single and double events (PATTERN $<= 4$), excluding low pulse-height events and hot pixel. Then we perform GTI selection in order to exclude background flare and extract source spectra from 0$^\prime$.2 radius of target centroid, using \verb'XSELECT'. We extract background spectra from outer region of a circle of 2$^\prime$.90 radius excluding verge pixels of the CCD chip. We merged them together and binned with the same criterion. The data have a 392.7 ks exposure after the GTI selection. We fit the spectrum with the same model. Figure \ref{fig:epn_sum} shows the spectrum and best fit model with $\chi^2_r = 3.06$ for 147 dof, where we confirm the keV-excess.

\section{Possible cause of the keV-excess}

We quantify the keV-excess and consider its possible causes, i.e., uncertainty in background, pile up of the low energy photons and confusion of other sources in the source region.

\subsection{Evaluation of the keV-excess and background}

We define the keV-excess fraction, $f_{ex}$, in the 0.8--1.2\,keV band as 

\begin{equation}
f_{ex}=\frac{ c_{obs} - c_{mod} }{c_{mod}},
\label{eq:f_ex}
\end{equation}

\noindent where $c_{obs}$ is the observed (background subtracted) count rate and $c_{mod}$ is the count rate expected from the 2T blackbody model. As shown in table \ref{tab:evaluate_ex}, the keV-excess fraction $f_{ex}$ is $17.1\pm6.6$\% for XIS1, $25.6\pm7.3$\% for XIS0+3 and $16.0\pm2.1$\% for EPIC-pn, respectively. The values of $f_{ex}$ are statistically positive with confidence level of 99.96\% or higher.   In table \ref{tab:evaluate_ex} we also show  $c_{obs}$, $c_{mod}$ and background count rate in 0.8--1.2\,keV, $b$, for each instrument.  Note that $b$ is the background count rate normalized for the source extraction region and the vignetting-corrected.

\subsection{Systematic uncertainty in the background}
The keV-excess might be caused by systematic uncertainty in the background subtraction. We define the ratio of the background uncertainty, $\Delta b$, to the model count rate in 0.8--1.2\,keV band, $f_{bg}$, as
\begin{equation}
f_{bg} = \frac{\Delta b}{c_{mod}} .
\label{eq:beta}
\end{equation}

\noindent We examine the spatial fluctuation of the background using the same data used to be extracted from the source spectra. For XIS1 we merge all the observations of XIS1 and derive 0.8--1.2\,keV count rates in the six independent circular regions with 2$^\prime$.17 radius (the same radius we used for the extraction of the source spectra) each centred at 4$^\prime$.35 from J1856. We calculate the mean and standard deviation of these count rates and derive the spatial fluctuation of  $2.5\times10^{-4}$ s$^{-1}$ from the standard deviation by subtracting the statistical error in quadratic manner. This corresponds to the fractional uncertainty in the background, $\Delta b/b = 5.32$\% for XIS1. We repeat the same procedure for the XIS0+3 data and obtain $\Delta b/b$ of  7.49\%. We examine six circular regions with 0$^\prime$.2 radius each centred at 3$^\prime$.1 from the source with the EPIC-pn data. We then obtain $\Delta b/b$ of 37.6\% for EPIC-pn. Hence, the uncertainty of the keV-excess owing to the background subtraction, $f_{bg}$, is 6.6\%, 6.7\%, and 0.44\% for XIS1, XIS0+3, and EPIC-pn, respectively. The systematic uncertainty in the background subtraction is negligibly small for EPIC-pn, while the positive keV-excess is at 2 and 3 sigma level for XIS1 and XIS0+3, respectively.

\subsection{Possible pile up}

XIS and EPIC-pn have 8 s and 6 ms exposure per frame with normal mode and small-window mode, respectively. We have to examine a possible pile up, i.e., two or more photons coming onto one pixel during one frame time, which could make a hard tail in the spectra.  
We can calculate the probability that photons with energy $< 0.8$\,keV pile up to 0.8--1.2\,keV. The piles up count rate $c$ is calculated as

\begin{equation}
 c = \lambda \int \int f(\epsilon^\prime-\epsilon)f(\epsilon^\prime) d\epsilon^\prime d\epsilon
\label{equation:pileup}
\end{equation}

\noindent where 

\begin{equation}
\lambda = \frac{\tau}{n_p}.
\end{equation}

\noindent $\epsilon$ is the photon energy, $f(\epsilon)$ is the spectrum of the photons within the source extract region in the unit of counts s$^{-1}$ keV$^{-1}$, assuming the 2T blackbody model. $\tau$ is the frame time of each CCD and $n_p$ is the number of CCD pixels within the source extraction region. The integration is done over the energy range which satisfies $0.8 < \epsilon < 1.2$ keV. Table \ref{tab:pileup} shows that the pile up count rate $c$ is $(2.23 \pm 0.05) \times10^{-6}$ s$^{-1}$ for XIS1, $(5.78 \pm 0.21) \times10^{-8}$ s$^{-1}$ for XIS0+3 and $(4.56 \pm 0.03) \times10^{-4}$ s$^{-1}$ for EPIC-pn, respectively. Thus the excess fraction from the pile up $c/c_{mod}$ is 0.07\% for XIS1, 0.003\% for XIS0+3 and 2.5\% for EPIC-pn, respectively. These values are significantly smaller than the  observed keV-excess, 16$\sim$25\%; the pile up hardly explain the keV-excess. 

\subsection{Possible confusing sources around J1856}

As we take relatively large extraction region of 2$^\prime$.17 radius circle for the XIS, we have to check confusion of other sources to the source spectra. For that purpose, we employ the XMM-Newton pipeline products of EPIC-MOS 2 data observing J1856 with large-window mode or full-frame mode (table \ref{tab:xmmdata_mos}). We search for sources within 2$^\prime$.17 radius of the target centroid and list them up (table \ref{tab:mos_sources}). Note that no sources is included in the 0$^\prime$.2 radius circle, the source extraction region of EPIC-pn. There is no sources detected in any of two observations out of twelve. We thus estimate the total count rate from them. The exposure time weighted average of that in 0.8--1.2\,keV is $(3.12\pm1.58) \times 10^{-4}$ s$^{-1}$. Converting this EPIC-MOS count rate to that of Suzaku by using \verb'PIMMS' and assuming a power-law spectrum with $\Gamma = 2$ leads to $(2.0\pm1.0) \times 10^{-4}$ s$^{-1}$ for XIS1 and $(1.2\pm0.6) \times 10^{-4}$ s$^{-1}$ for XIS0+3. Dividing this by the 0.8--1.2\,keV count rate of the 2T blackbody model $c_{mod}$, we obtain the excess fraction from the confusing sources of $6.2\pm3.2$\% for XIS1 and $7.0\pm3.5$\% for XIS0+3. This is smaller than the observed keV-excess, 17\% and 26\% for XIS1 and XIS0+3, respectively.  As we have not explicitly excluded sources as faint as those listed in table 8 in our XIS analysis, we can consider that the contribution of the confusing sources are already taken into account in the uncertainty of the background. 

\begin{figure}
 \begin{center}
  \includegraphics[width=10cm]{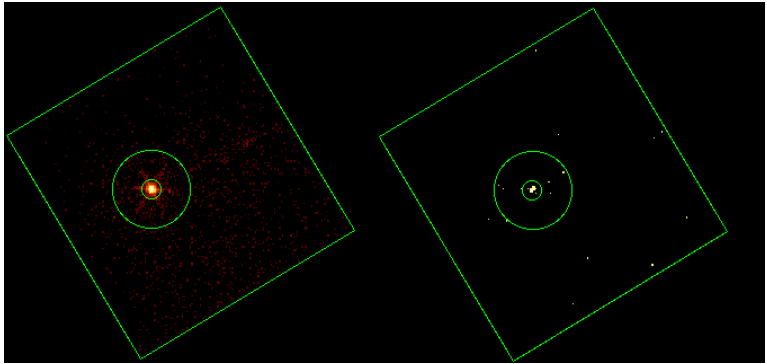} 
 \end{center}
\caption{X-ray images obtained with Chandra, selecting 0.2--1.2\,keV ({\sl left}) and 0.8--1.2\,keV ({\sl right}) energy band. The squares correspond to $1^\prime.27 \times 1^\prime.33$, while the inner and outer circles correspond to 0$^\prime$.05 and 0$^\prime$.2 radius from J1856, respectively. No other sources than J1856 is apparent in these images.}\label{fig:chandra}
\end{figure}

On the other hand, we are still not able to rule out the possibility that any sources within 0$^\prime$.2 from J1856 might be the origin. We thus investigate the X-ray images with Chandra data obtained in 2011-06-10 with 28.4 ks exposure (Obs ID: 13198). We show the 0.2--1.2\,keV image and 0.8--1.2\,keV image in figure \ref{fig:chandra}. We regard 30 events within 0$^\prime$.05 radius (inner circle in figure \ref{fig:chandra}) as intrinsic emission from J1856. We find extra 6 events between 0$^\prime$.05--0$^\prime$.2 (outer circle). Four of them are spoke-like component of PSF by mirror support struts and 2 of them are out-of-time events. Since the background is expected to be 1 counts for 0$^\prime$.05-0$^\prime$.2, we obtain the excess fraction from the confusing sources within 0$^\prime$.2, $-3.3\pm5.7$\%.

\section{Spectral fitting for the keV-excess}

\begin{figure}
 \begin{center}
  \includegraphics[width=10cm]{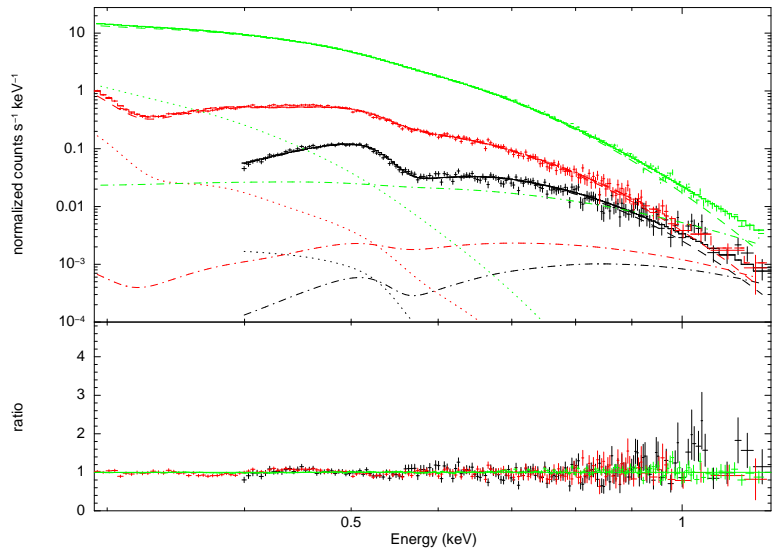} 
 \end{center}
\caption{Fit with three-temperature blackbody model. Parameters for the first and second components are fixed to the best fit values of Beuermann et al. (2006), while those of the third component are obtainedby the fit as $kT^\infty = 137$ keV and $R^\infty = 36$ m. The dash-doted lines show the third blackbody component. Black is XIS0+3 (FI), red is XIS1 and green is EPIC-pn, respectively..}\label{fig:3bb}
\end{figure}

\begin{figure}
 \begin{center}
  \includegraphics[width=10cm]{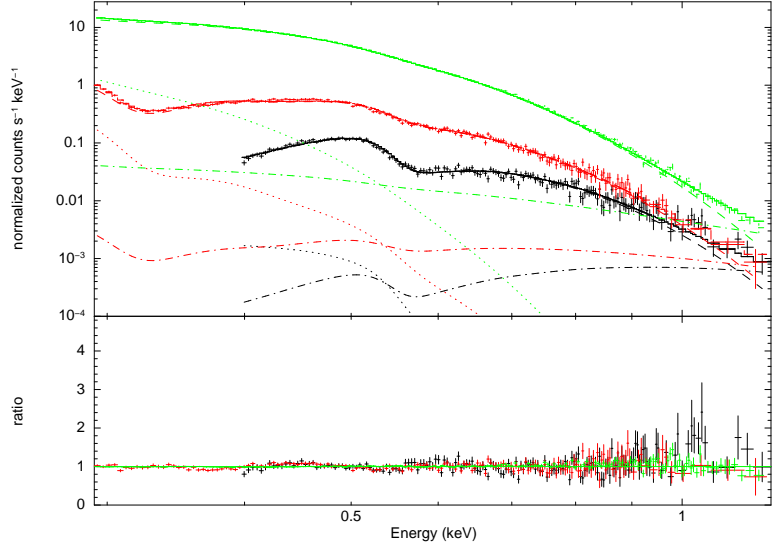} 
 \end{center}
\caption{Fit with the fixed 2T blackbody and a power-law model. The best fit gives $\chi^2_r = 1.51$ for 540 dof with photon index $\Gamma = 3.36$ and normalization $n = 5.2 \times 10^{-6}$. The dash-doted lines show the power-law component. Color index is the same with figure \ref{fig:3bb}.}\label{fig:pl}
\end{figure}

\begin{figure}
 \begin{center}
  \includegraphics[width=10cm]{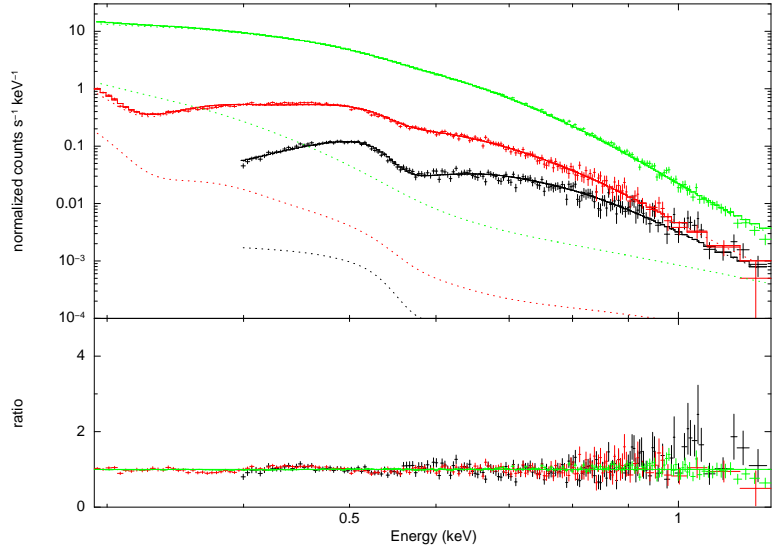} 
 \end{center}
\caption{Fit with the fixed 2T blackbody, convoluted with `simpl', a model of Comptonization of seed spectrum into power-law. The best fit gives $\chi^2_r = 1.46$ for 540 dof with photon index $\Gamma = 4.3$ and the scattering fraction $f_{sc} = 1.9 \times 10^{-3}$.  Color index is the same with figure \ref{fig:3bb}.}\label{fig:comp}
\end{figure}

We have shown that the keV-excess in J1856 observed with three instruments is not explained  by uncertainty in the background subtraction, pile up, nor confusing sources. We then fit the keV-excess spectra with a blackbody or power-law model as an additional component to the baseline 2T blackbody model. When we use a blackbody model for the keV-excess, the best fit parameters are $kT^\infty = 137^{+18}_{-14}$ eV and $R^\infty = 36^{+45}_{-36}$ m with $\chi^2_r = 1.42$ for 540 dof. If we employ a power-law model for the keV-excess, we obtain $\Gamma = 3.4^{+0.5}_{-0.6}$ and normalization at 1 keV $n=(5.2^{+0.9}_{-0.9}) \times 10^{-6}$ keV$^{-1}$ cm$^{-2}$ s$^{-1}$ with $\chi^2_r = 1.51$ for 540 dof. Moreover, we apply a convolution model of Comptonization, simpl, which describes a scenario that blackbody seed photons are repeatedly scattered by hot electrons into power-law (Steiner et al. 2009). We consider only up-scattering and obtain the best fit parameters, photon index $\Gamma_c = 4.3^{+0.8}_{-0.8}$ and the scattered fraction $f_{sc} = (1.9^{+0.9}_{-1.1}) \times 10^{-3}$ with $\chi^2_r = 1.46$ for 540 dof. The spectra with these best fit models are shown in figure \ref{fig:3bb}, figure \ref{fig:pl} and figure \ref{fig:comp}, respectively. Assuming the target distance of 120 pc, the luminosity of the keV-excess in the 0.4--1.2 keV band $L_{ex}$ is $\sim 4 \times 10^{28}$ erg s$^{-1}$ and $\sim 3 \times 10^{28}$ erg s$^{-1}$ for the blackbody and the power-law, respectively, while the bolometric luminosity of the cool and hot component in the baseline 2T blackbody model is $4.3 \times 10^{-31}$ erg s$^{-1}$ and  $4.7 \times 10^{-31}$ erg s$^{-1}$, respectively. 

Either model reduces the $\chi^2$ values from the original 2T blackbody model fit ($\chi^2_r=1.93$ for 542 dof). The improvement of the fit is statistically examined with F-test.  F values is 97.5 for the additional blackbody component, 76.8 for the additional power-law component and 86.0 for the Comptonization model. Either case is highly significant with  significance level of $10^{-29}$ or less.

\section{Search for pulsation in the keV-excess}

\begin{figure}
 \begin{center}
  \includegraphics[width=10cm]{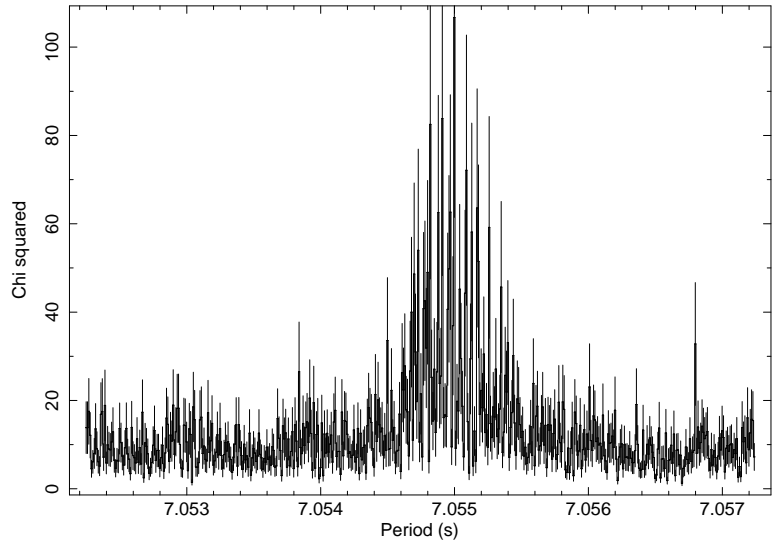} 
 \end{center}
\caption{Folding period versus $\chi^2$, using the EPIC-pn data in the 0.2--1.2\,keV band.} \label{fig:search}
\end{figure}

\begin{figure}
 \begin{center}
  \includegraphics[width=10cm]{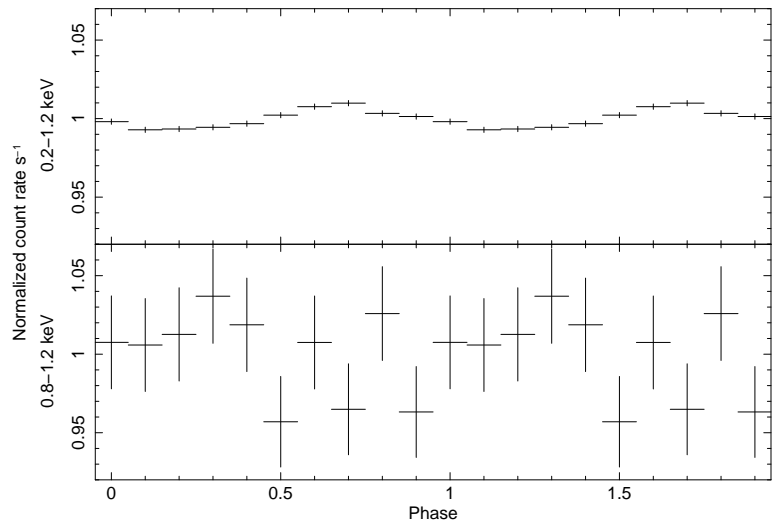} 
 \end{center}
\caption{Folded light curves of the EPIC-pn data with 7.05525 s in the 0.2--1.2\,keV band ({\sl above}) and the 0.8--1.2\,keV band ({\sl below}).}\label{fig:lc}
\end{figure}

Tiengo and Mereghetti (2007) reported the pulsation of J1856 with 1.5\% pulsed fraction and 7.055 s period in 0.15--1.2\,keV energy range with the EPIC data. It is interesting to search for a possible pulsation in the energy range including the keV-excess. We thus perform a coherent timing analysis of the EPIC-pn event data (table \ref{tab:xmmdata}) in 0.8--1.2\,keV band within the 0$^\prime$.2 radius circle. We employ \verb'efsearch' program for the epoch folding analysis to search for a period in 4--11\,s, as was done by Tiengo and Mereghetti (2007) with time interval of $10^{-4}$ s. However, we obtain no significant peak with $\chi^2 > 50$ for 9 dof. We then search for periods of the data coherently in 0.2-1.2 keV in 4--11\,s and obtain 7.05525 s with $\chi^2$ of 138.3. Figure \ref{fig:search} shows the periodgram. We fold the 0.8--1.2\,keV data with this period to derive the pulsed fraction for the data by fitting sinusoid to the folded light curve, shown in figure \ref{fig:lc}. We obtain the upper limit of the pulsed fraction of 3\% for 90\% confidence level.

\section{Discussion}

We have discovered the keV-excess in the X-ray spectra of J1856 observed with three independent instruments. We examined possible causes of the keV-excess, i.e., uncertainty in the background subtraction, the pile up effect, and the contribution of the confusing sources. None of them accounts for the keV-excess. We thus consider that the keV-excess is most likely intrinsic to J1856. 

If we adopt the blackbody model for the keV-excess, the temperature $kT^\infty$ is 137 eV and the radiation radius $R^\infty$ is 36 m, indicating the emission of a small area. In the baseline 2T blackbody model, the hot component ($kT^\infty_h \sim 62$ eV, $R^\infty_h \sim 5$ km) is usually considered as a hot spot on the surface of the neutron star. The keV-excess might be a super hot spot, but we do not know any theoretical model to support such a possibility. 

Ho et al. (2007) suggested that their partially ionized hydrogen atmosphere model ($B = 4 \times 10^{12}$ G in Table 1) could also reproduce the spectrum of J1856 as well as the baseline 2T blackbody model. Their model predicts 17\% lower flux at 1 keV than that of the 2T blackbody model if we normalize these two models at 0.8 keV. It means that their partially ionized hydrogen atmosphere model cannot be the origin of the keV-excess.

Alternatively when we adopt the power-law model, the keV-excess could be a non-thermal emission. Synchrotron radiation by relativistic electrons is a possibility to be considered, which is characterized with the critical frequency $\omega_c$;

\begin{equation}
 \omega_c \sim \frac{eB}{m_ec}\gamma^2
\label{equation:synchro}
\end{equation}

\noindent where $\gamma$ is Lorentz factor of electrons and $B$ is the magnetic field. The non-thermal radiation from the pulsars are observed with a photon index $\Gamma \sim 1.5$, which is inconsistent with that of J1856, $\Gamma \sim 3.4$. This soft and steep spectrum indicates that the energy peak of synchrotron radiation is below 1 keV. If the peak is 0.01 keV, we obtain $B \sim 10^{13}\gamma^{-2}$ G.

Comptonization of the blackbody photons is another possibility as we explicitly employ `simpl' model in section 4. We would note an interesting analogy between the keV-excess in J1856 and the ``soft tail'' component in the soft X-ray spectrum of magnetars. Magnetars are isolated neutron stars with strong magnetic field, $B \sim 10^{14} - 10^{15}$ G. Suzaku observations revealed that their spectra consist of soft blackbody ($< 10$ keV energy range) and a hard power-law ($> 10$ keV; Enoto et al. 2010b). However, Enoto et al. (2010a and 2011) showed that the soft X-ray spectra of magnetars have a ``soft tail'' above a blackbody model around 10 keV toward higher energies. They stated that Comptonized blackbody model or Resonant Cyclotron Scattering model (RCS; Rea et al. 2008) with blackbody seed photons can reproduce the soft tail with photon index $\Gamma \sim 4$, which is consistent with our result of Comptonized blackbody model ($\Gamma_c = 4.3$). We should also try to fit the keV-excess with RCS model, however, the publically released RCS model only allows the blackbody temperature of seed photons $> 0.1$ keV, which is higher than J1856's temperature, $kT^\infty \sim 32$ or $63$ eV. Considering that the magnetic field of J1856 is one or two orders of magnitude weaker than those of magnetars, the keV-excess might be a weak version of the soft tail component in magnetars.

We can rule out neither emission model at this point. We apparently need further observations of the keV-excess to approach its origin. As we employed hundreds ks data of Suzaku and XMM-Newton, we may need instruments with much larger effective area. NICER (\cite{gendreau_2012})  and ATHENA (\cite{barret_2013}) might be candidates for that purpose.

\section{Summary}
Fitting the spectra of RX J1856.5$-$3754 obtained with three independent instruments, Suzaku XIS1, XIS0+3 and XMM-Newton EPIC-pn with the known 2T blackbody model, we find that the data similarly exceed the model significantly in the $>0.8$ keV energy range. This keV-excess is evaluated with the excess fraction, 17.1$\pm$7.8\% for XIS1, 25.6$\pm$7.0\% for XIS0+3 and 16.0$\pm$2.2\% for EPIC-pn. We then consider possible causes of the keV-excess, uncertainty in background, pile up of the low energy photons and confusion of other sources in the source region. However, none of them can explain the keV-excess. We thus consider the keV-excess is most likely intrinsic to J1856. We then fit the keV-excess spectra with a blackbody or power-law model as an additional component to the baseline 2T blackbody model. When we use a blackbody model, the best fit temperature $kT^\infty$ is $137^{+18}_{-14}$ eV, while we obtain $\Gamma = 3.4^{+0.5}_{-0.6}$ employing a power-law model. Furthermore, we apply a convolution model of Comptonization with blackbody seed photons and obtain $\Gamma_c = 4.3^{+0.8}_{-0.8}$.

On the other hand, J1856 has a 7.055 s period pulsation with 1.5\% pulsed fraction in 0.15--1.2\,keV (Tiengo \& Mereghetti 2007). We confirm it by using our data coherently in 0.8--1.2\,keV. We thus search for a pulsation in the 0.8--1.2\,keV band, but we find no significant pulsation in the period of 4--11\,s. We obtain the upper limit of the pulsed fraction of $3$\% for the pulse period of 7.05525 s.

We consider three emission models for the origin of the keV-excess, while we can rule out neither of them. However, we would note an interesting analogy between the keV-excess in J1856, of which magnetic field $B \sim 10^{13}$ G and the soft tail ($>$10 keV) component of the magnetars that have $B\sim10^{14}-10^{15}$  G. Considering the magnetic field of J1856 is one or two order of magnitude weaker than those of magnetar, the keV-excess might be a weak version of the soft tail component in magnetars.



\begin{ack}
We are very grateful to Wynn G. C. Ho for kindly providing their table model. We thank Teruaki Enoto for his advice on timing analysis. This work is supported by Japan Society for the Promotion of Science (JSPS) KAKENHI Grant Number 16H00949, 26109506, 16K13787, 23000004, 15J018450, 24684010, 26670560, 15H03641.

\end{ack}






\newpage


\newpage



\begin{table}[htbp]
  \tbl{XIS observations of J1856}{%
  \begin{tabular}{cccc}
\hline  
Obs. ID&Date&Exposure [ks] \\ \hline \hline
100041010&2005-10-24&76.25 \\ 
101009010&2006-10-20&40.78 \\ 
102014010&2007-10-15&41.32 \\ 
103006010&2008-10-20&43.05 \\ 
104022010&2009-10-23&43.49 \\ 
104022030&2010-03-26&42.45 \\ 
106009010&2011-10-22&39.32 \\ 
107007020&2012-10-20&44.05 \\ 
108007010&2013-04-15&40.75 \\ 
109008010&2014-04-08&38.97 \\ \hline
Total&-&450.43 \\ \hline
    \end{tabular}}\label{tab:suzaku_data}
\begin{tabnote}

\end{tabnote}
\end{table}

\begin{table}[htbp]
  \tbl{EPIC-pn data of J1856}{%
  \begin{tabular}{cccc}
\hline  
Obs. ID&Date&Exposure [ks] \\ \hline \hline
0165971601&2004-09-24&23.10 \\
0165972001&2005-12-29&23.26 \\
0165972101&2006-03-26&48.69 \\
0412600301&2007-10-04&15.24 \\
0412600601&2008-10-23&43.79 \\
0412600801&2009-10-07&43.79 \\
0412601101&2010-09-28&48.22 \\
0412601501&2011-10-05&17.49 \\
0412602301&2012-09-20&56.16 \\
0727760101&2013-09-14&49.68 \\
0727760301&2014-09-18&55.01 \\
0727760501&2015-10-03&57.08 \\ \hline
    \end{tabular}}\label{tab:xmmdata}
\begin{tabnote}
All observations are performed with small-window mode for pn.
\end{tabnote}
\end{table}

\begin{table}[htbp]
  \tbl{Count rates in 0.8--1.2\,keV}{%
  \begin{tabular}{cccc}
\hline  
Count rate [10$^{-3}$s$^{-1}]$&XIS1&XIS0+3&EPIC-pn\\ \hline \hline
$c_{obs}$&3.79$\pm$0.23&2.24$\pm$0.12&20.7$\pm$0.3 \\
$c_{mod}$&3.23&1.78&17.9 \\ 
$b$&4.70$\pm$0.30&2.00$\pm$0.99&0.250$\pm$0.005 \\ \hline
$f_{ex}$ [\%]&17.1$\pm$6.6&25.6$\pm$7.3&16.0$\pm$2.1 \\ \hline
    \end{tabular}}\label{tab:evaluate_ex}
\begin{tabnote}
\end{tabnote}
\end{table}

\begin{table}[htbp]
  \tbl{Estimation of pile up count rate in $>0.8$ keV band}{%
  \begin{tabular}{cccc}
\hline  
Instrument& $\lambda$ [$10^{-4}$ s] & $c$ [$10^{-6}$ s$^{-1}$] & $\frac{c}{c_{mod}}$ [\%] \\ \hline \hline
XIS1 & 1.63 & 2.23$\pm$0.05 & 0.070$\pm$0.003 \\
XIS0+3 & 1.63 & 0.058$\pm$0.002 & 0.0032$\pm$0.0001 \\
EPIC-pn &  2.23 & 456$\pm$3 & $2.53\pm0.02$ \\
\hline
    \end{tabular}}\label{tab:pileup}
\begin{tabnote}
\end{tabnote}
\end{table}

\begin{table}[htbp]
  \tbl{EPIC-MOS 2 data of J1856 in full frame mode or large window mode}{%
  \begin{tabular}{ccccc}
\hline  
Obs. ID&Date&Exposure [ks] &Mode\\ \hline \hline
0201590101&2004-04-17&66.45&FF \\
0415180101&2007-03-25&40.91&LW \\
0412601101&2010-09-28&69.93&LW \\
0412601301&2011-03-14&83.43&LW \\
0412601501&2011-10-05&17.49&LW \\
0412601401&2012-04-13&77.11&LW \\
0412602201&2013-03-14&73.91&LW \\
0727760201&2014-03-26&75.00&LW \\
0727760401&2015-03-12&75.60&LW \\ 
0727760601&2016-03-11&78.00&LW \\ \hline
    \end{tabular}}\label{tab:xmmdata_mos}
\begin{tabnote}
FF indicates full-frame mode and LW does large-window mode.
\end{tabnote}
\end{table}

\begin{table}[htbp]
  \tbl{Source detected within 2$^\prime$.17 radius source region by EPIC-MOS 2}{%
  \begin{tabular}{cccccc}
\hline  
Date&R.A [h m s]&Dec [$^\circ$ $^\prime$ $^{\prime\prime}$]&$\phi$ [$^\prime$]&\multicolumn{2}{c}{Count rate} \\ 
&&&&0.2--12\,keV [$10^{-2}$ s$^{-1}$]&0.8--1.2\,keV [$10^{-4}$ s$^{-1}$] \\ \hline \hline
2004-04-17&18 56 39.60&$-$37 54 52.1&0.82&2.30$\pm$0.51&1.30$\pm2.11$ \\
2007-03-25&-&-&-&-&- \\
2010-09-28&18 56 33.53&$-$37 55 52.3&1.34&2.35$\pm$0.43&0.70$\pm1.13$ \\
2011-03-14&18 56 33.94&$-$37 53 32.1&1.15&6.10$\pm$0.45&6.34$\pm2.18$ \\
2011-10-05&18 56 36.49&$-$37 55 40.1&1.06&2.65$\pm$0.45&3.00$\pm1.30$ \\
2012-04-13&-&-&-&-&- \\
2013-03-14&18 56 32.05&$-$37 55 38.6&1.27&2.19$\pm$0.45&3.33$\pm1.48$ \\
&18 56 41.89&$-$37 54 46.8&1.19&2.01$\pm$0.38&45.5$\pm4.61$ \\
2014-03-26&18 56 35.10&$-$37 53 34.0&1.08&2.68$\pm$0.54&2.43$\pm1.68$ \\
2015-03-12&-&-&-&-&- \\
2016-03-11&18 56 41.10&$-$37 55 2.1&1.05&2.01$\pm$0.74&2.72$\pm1.79$ \\
&18 56 35.92&$-$37 53 18.1&1.33&1.53$\pm$0.26&3.11$\pm1.79$ \\ \hline
Average&&&&&3.12$\pm1.58$\\ \hline
    \end{tabular}}\label{tab:mos_sources}
\begin{tabnote}
$\phi$ is the angular distance of the sources from J1856. Background-subtracted count rates are shown.
\end{tabnote}
\end{table}


\end{document}